\documentclass[twocolumn, trackchanges]{aastex7}

\usepackage{graphicx}	
\usepackage{amsmath}	
\usepackage{svg}
\usepackage{xspace}
\usepackage{xcolor}
\usepackage{orcidlink}
\usepackage{soul}

\providecommand{\bjdtdb}{\ensuremath{\rm {BJD_{TDB}}}}
\providecommand{\tjdtdb}{\ensuremath{\rm {TJD_{TDB}}}}

\providecommand{\teff}{\ensuremath{T_{\rm eff}}}

\providecommand{\msun}{\ensuremath{\,M_\Sun}}
\providecommand{\rsun}{\ensuremath{\,R_\Sun}}
\providecommand{\lsun}{\ensuremath{\,L_\Sun}}
\providecommand{\mj}{\ensuremath{\,M_{\rm J}}}
\providecommand{\rj}{\ensuremath{\,R_{\rm J}}}

\providecommand{\tess}{{TESS\,}}
\providecommand{\meterps}{{$\rm m\, \rm s^{-1}$}}

\graphicspath{{./}{figures/}}

\providecommand{\dbf}{}

\begin{document}

\title{The OATMEAL Survey. II. The 3D spin-orbit obliquity of an eccentric transiting brown dwarf in the Ruprecht 147 open cluster}

\correspondingauthor{Theron W. Carmichael}
\email{tcarmich@hawaii.edu}
\author[0000-0001-6416-1274, sname=Carmichael, gname=Theron]{Theron W. Carmichael}
\altaffiliation{NSF Ascend Postdoctoral Fellow}
\affiliation{Institute for Astronomy, University of Hawai‘i, 2680 Woodlawn Drive, Honolulu, HI 96822, USA}
\email{tcarmich@hawaii.edu}

\author[0000-0002-8965-3969, sname=Giacalone, gname=Steven]{Steven Giacalone}
\altaffiliation{NSF Astronomy and Astrophysics Postdoctoral Fellow}
\affiliation{Department of Astronomy, California Institute of Technology, Pasadena, CA 91125, USA}
\email{giacalone@astro.caltech.edu}

\author[0000-0002-0701-4005, sname=Vowell, gname=Noah]{Noah Vowell}
\affiliation{Center for Data Intensive and Time Domain Astronomy, Department of Physics and Astronomy, Michigan State University, East Lansing, MI 48824, USA}
\affiliation{Center for Astrophysics \text{\textbar} Harvard \& Smithsonian, 60 Garden St, Cambridge, MA 02138, USA}
\email{vowellno@msu.edu}

\author[0000-0001-8832-4488, sname=Huber, gname=Daniel]{Daniel Huber}
\affiliation{Institute for Astronomy, University of Hawai‘i, 2680 Woodlawn Drive, Honolulu, HI 96822, USA}
\email{huberd@hawaii.edu}

\author[0000-0002-0376-6365, sname=Wang, gname=Xian-Yu]{Xian-Yu Wang}
\affiliation{Department of Astronomy, Indiana University, 727 East 3rd Street, Bloomington, IN 47405, USA}
\email{xwa5@iu.edu}

\author[sname=Bossett, gname=Malik]{Malik Bossett}
\affiliation{Department of Astronomy and Astrophysics, University of California, Santa Cruz, CA 95064, USA}
\email{mbossett@ucsc.edu}

\author[0000-0002-9305-5101, sname=Handley, gname=Luke]{Luke Handley}
\affiliation{Department of Astronomy, California Institute of Technology, Pasadena, CA 91125, USA}
\email{lhandley@caltech.edu}

\author[0000-0002-5812-3236, sname=Householder, gname=Aaron]{Aaron Householder}
\affiliation{Department of Earth, Atmospheric and Planetary Sciences, Massachusetts Institute of Technology, Cambridge, MA 02139, USA}
\affil{Kavli Institute for Astrophysics and Space Research, Massachusetts Institute of Technology, Cambridge, MA 02139, USA}
\email{aaron593@mit.edu}

\author[0000-0003-3020-4437, sname=Li, gname=Yaguang]{Yaguang Li}
\affiliation{Institute for Astronomy, University of Hawai‘i, 2680 Woodlawn Drive, Honolulu, HI 96822, USA}
\email{yaguangl@hawaii.edu}

\author[0000-0003-3504-5316, sname=Fulton, gname=Benjamin]{Benjamin J. Fulton}
\affiliation{Cahill Center for Astronomy \& Astrophysics, California Institute of Technology, Pasadena, CA 91125, USA}
\affiliation{IPAC-NASA Exoplanet Science Institute, Pasadena, CA 91125, USA}
\email{bjfulton@ipac.caltech.edu}

\author[0000-0001-8638-0320, sname=Howard, gname=Andrew]{Andrew W. Howard}
\affiliation{Cahill Center for Astronomy \& Astrophysics, California Institute of Technology, Pasadena, CA 91125, USA}
\email{ahoward@caltech.edu}

\author[0000-0002-0531-1073, sname=Isaacson, gname=Howard]{Howard Isaacson}
\affiliation{Department of Astronomy, University of California Berkeley, Berkeley, CA 94720, USA}
\email{hisaacson@berkeley.edu}

\author[0000-0003-1312-9391, sname=Halverson, gname=Samuel]{Samuel Halverson}
\affiliation{Jet Propulsion Laboratory, California Institute of Technology, 4800 Oak Grove Drive, Pasadena, CA 91109, USA}
\email{samuel.halverson@jpl.nasa.gov}

\author[0000-0001-8127-5775, sname=Roy, gname=Arpita]{Arpita Roy}
\affiliation{Astrophysics \& Space Institute, Schmidt Sciences, New York, NY 10011, USA}
\email{arpita308@gmail.com}



\begin{abstract}

We present new analysis of the CWW 89 system as part of the Orbital Architectures of Transiting Massive Exoplanets And Low-mass stars (OATMEAL) survey. The CWW 89 system is a member of the 2.8 Gyr old Ruprecht 147 (NGC 6774) cluster and consists of two stars, CWW 89A (EPIC 219388192) and CWW 89B, with the primary hosting a transiting brown dwarf, CWW 89Ab. We use in-transit, highly precise radial velocity measurements with the Keck Planet Finder (KPF) to characterize the Rossiter-McLaughlin (RM) effect and measure the projected spin-orbit obliquity $|\lambda|=1.4\pm2.5^\circ$ and the full 3D spin-orbit obliquity of the brown dwarf to be $\psi=15.1^{+15.0^\circ}_{-10.9}$. This value of $\lambda$ implies that the brown dwarf's orbit is prograde and well-aligned with the equator of the host star, continuing the trend of transiting brown dwarfs showing a preference for spin-orbit alignment ($\lambda \approx 0^\circ$) regardless of the stellar effective temperature. This contrast with the transiting giant planet population, whose spin-orbit alignments depend on host $\teff$, showing an increasingly clear distinction in the formation and orbital migration mechanisms between transiting giant planets and transiting brown dwarfs like CWW 89Ab. For this system in particular, we find it plausible that the brown dwarf may have undergone coplanar high-eccentricity migration influenced by CWW 89B.

\end{abstract}


\section{Introduction} \label{sec:intro}

Our understanding of the tidal realignment of close-in planetary orbits around main sequence stars has substantially improved since the first measurements of spin-orbit obliquities were made for gas giant planets \citep{first_rm, winn2005, triaud2010, hirano2011, albrecht2012}. Here, we define ``spin-orbit obliquity'' to mean the angle between the rotation axis of the primary star and the orbital axis of the companion. With recent works by \cite{zak2024, espinoza2024, knudstrup2024, wang2024, jingwen2025}, the number of obliquity measurements made for giant planets now numbers in the dozens, with a wide range of aligned to misaligned systems. A relatively coherent story is emerging for the orbital evolution of these systems and it is centered on a general theory of the interior structures of stars that host giant planets. As main sequence stars that are more massive than the Sun evolve, expand in radius, and decrease in effective temperature, their outer envelopes transition from being predominantly radiative to convective. \cite{kraft1967} first established a temperature threshold at which this change occurs, now known as the Kraft Break at a $\teff = 6250$K. From this, a trend in the projected spin-orbit alignments, or obliquities $\lambda$, for giant planets around the Kraft Break has emerged: Stars hotter than the Kraft Break show a population of planets with no preferential value of $\lambda$ whereas stars cooler than the Kraft Break yield a companion planet population with preferentially aligned obliquities ($\lambda = 0^\circ$) \cite[e.g.][]{winn2010, rice2022, espinoza2023_CHEM, rusznak2024}. Recent observations of subgiant stars that have crossed the Kraft break during their evolution confirmed that alignments of giant planets are strongly tied to presence of convective envelopes \citep[e.g.][]{saunders24}.

The more massive siblings to giant planets, brown dwarfs (BDs), have recently seen a boom in their known transiting population in large part thanks to the \tess mission \citep{tess} and efforts from ground-based facilities like the Next Generation Transit Survey (NGTS) \citep{ngts}. Recent works from \citep[][and references therein]{henderson2024, vowell2025, larsen2025} show more than 50 known transiting brown dwarfs and like with their giant planet siblings, astronomers have pursued measurements of their spin-orbit obliquities for insight into their orbital histories. However, in contrast to the dozens spin-orbit alignments measured for transiting giant planets, only 9 such measurements exist for transiting BDs. Every measurement of $\lambda$ for transiting BDs thus far indicates a preference for an aligned orbit \citep{xo3b_1, corot3_RM, kelt1b, wasp30_rm, hats70b, oatmeal1, dosSantos2024, brady2025, doyle2025}. 

One widely-used technique to measure $\lambda$ for  transiting system is the Rossiter-McLaughlin (RM) Effect \citep{first_rm, rm_effect}. This requires taking spectra of the host star during a transit to observe the radial velocity (RV) anomaly induced by the transiting companion blocking varying amounts of red- or blue-shifted light depending on the orbital configuration relative to the orientation of the host star's spin axis. The RM effect alone only offers an estimate of the \textit{projected} spin-orbit alignment $\lambda$, but in some cases, the system supplies all of the ingredients needed to determine the full 3D obliquity $\psi$, namely  the orbital inclination $i_0$ of the transiting companion and the stellar spin axis angle $i_\star$, which is determined from combining the projected rotational velocity of the star with its rotation period (often observed via photometric modulations in its light curve).

Here we present a Rossiter-McLaughlin measurement of $\lambda$ for the transiting BD CWW 89Ab \dbf{ (also known as EPIC 219388192b and TOI-7497)}. CWW 89Ab's environment distinguishes it from other BD systems with measured obliquities. The CWW 89 system receives its name from \cite{curtis2013}, who catalog it and over 100 other stars as members of the Ruprecht 147 open cluster. This makes CWW 89Ab one of four known transiting BDs that reside in a stellar association or cluster. The other three are AD 3116b in Praesepe, RIK 72b in Upper Scorpius, and HIP 330609b in MELANGE 6 from \cite{ad3116}, \cite{david19_bd}, and \cite{vowell2023}, respectively. CWW 89Ab is the oldest BD to reside in such an environment at $2.76\pm 0.61$ Gyr \citep{r147_age}. CWW 89 consists of a primary G-dwarf star, CWW 89A, and a secondary M-dwarf, CWW 89B, orbiting at a 24.9 AU (81 mas) projected separation \citep{cww89a}. A handful of transiting BD systems comprise 3 or more objects, typically a primary star, a transiting BD companion, and a more distant outer stellar companion, but CWW 89 is the only multi-star transiting BD system featuring a G-dwarf star as its primary. 

The paper proceeds as follows: Section \ref{sec:obs} details past and new RV measurements used in the orbital solution and characterization of the RM effect. Section \ref{sec:analysis} details our RM model, calculation of the stellar spin axis angle, and the determination of the full 3D obliquity. Section \ref{sec:discussion} outlines considerations for the emerging trend in spin-orbit alignments for the transiting brown dwarf population. \dbf{   Section \ref{sec:summary} recapitulates our most important findings.}

\section{Observations}\label{sec:obs}

\subsection{Keck Planet Finder in-transit RVs}
From 07:10 UT to 12:19 UT on 2024 July 30, we observed a transit of the CWW 89Ab using the Keck Planet Finder \cite[KPF;][]{Gibson2024} with the Keck I telescope at the W. M. Keck Observatory located on Maunakea, HI, USA. To ensure sufficient pre- and post-transit baseline, we observed the star continuously from approximately 1 hour before ingress to 1 hour after egress. We took a series of spectra with KPF at 600\,s exposure times each, resulting in 20 in-transit observations at a SNR of 60-90 across the 445-600nm wavelength range. The goal was to maintain a SNR at this level or higher while keeping good temporal sampling of the transit so that the RV anomaly induced by the RM effect would be well-detected. The RVs are calculated using the KPF data reduction pipeline \citep{KPF_DRP}. The complete series of KPF RVs are given in Table \ref{tab:kpf_rvs}. We note that KPF produces RVs in a green and red CCD, taking the RV values from different orders in the echelle spectrum \dbf{   with the green CCD spanning 445-600nm and the red CCD spanning 600-870nm}. We only use the RVs derived from the green CCD as these are at a higher SNR than those of the red CCD for this star by a factor of $\lesssim2.5$. 

Based on reported the stellar parameters in \cite{nowak17} and \cite{carmichael19} of $M_\star=1.1\rm\msun$, $R_\star=1.0\rm\rsun$, $\teff=5755$K, $\log{g}=4.5$, and $\rm [Fe/H]=+0.2$, we assume CWW 89A is a sunlike star.

\begin{deluxetable*}{ccc}
\tabletypesize{\footnotesize}
\tablewidth{0pt}
\tablecaption{In-transit KPF relative RV time series for CWW 89A. We subtract the mean of the RVs (47262\meterps) arbitrarily for the purpose of displaying the data in this table. \label{tab:kpf_rvs}}
 \tablehead{
 \colhead{$\rm BJD_{\rm TDB}$} & \colhead{RV - 47250 ($\rm m\, s^{-1}$)} & \colhead{$\sigma_{\rm RV}$ ($\rm m\, s^{-1}$)}}
\startdata
2460521.8046937  &  526.92  &  4.28 \\
2460521.8119592  &  489.96  &  4.42 \\
2460521.8198269  &  446.04  &  5.19 \\
2460521.8267481  &  412.76  &  4.81 \\
2460521.8355091  &  361.17  &  4.67 \\
2460521.8415548  &  341.3  &  6.27 \\
2460521.8503282  &  304.77  &  4.4 \\
2460521.8575215  &  284.0  &  3.55 \\
2460521.8646204  &  243.27  &  5.09 \\
2460521.8725758  &  209.82  &  4.2 \\
2460521.8799776  &  168.5  &  3.79 \\
2460521.8876566  &  130.2  &  4.21 \\
2460521.8963321  &  79.69  &  4.89 \\
2460521.9031678  &  38.74  &  3.77 \\
2460521.9108818  &  -6.41  &  3.88 \\
2460521.9181893  &  -50.28  &  3.62 \\
2460521.9252793  &  -90.72  &  3.99 \\
2460521.9333313  &  -133.15  &  4.36 \\
2460521.9405187  &  -175.64  &  4.66 \\
2460521.948442  &  -210.8  &  4.58 \\
2460521.9559914  &  -253.12  &  4.18 \\
2460521.9633564  &  -281.51  &  4.15 \\
2460521.9709421  &  -308.74  &  4.16 \\
2460521.9782791  &  -328.87  &  4.04 \\
2460521.9859417  &  -365.23  &  3.94 \\
2460521.993371  &  -403.39  &  4.12 \\
2460522.0004741  &  -436.12  &  4.69 \\
2460522.0083231  &  -479.83  &  6.05 \\
2460522.0161146  &  -513.33  &  5.3 \\
\enddata
\end{deluxetable*}

\subsection{Literature RVs from TRES and FIES} \label{sec:lit_rvs}
Past works by \cite{nowak17} and \cite{carmichael19} have published orbital solutions (i.e. out-of-transit spectra) for CWW 89A. These data are used to remove the Keplerian motion from the in-transit RVs newly collected with KPF. These data aid in the detailed analysis of the RM effect to account for the out-of-transit motion of the brown dwarf.

\cite{carmichael19} provide 18 RVs measured from spectra taken with the Tillinghast Reflector Echelle Spectrograph (TRES) located at the Fred Lawrence Whipple Observatory at Mt. Hopkins, AZ, USA. These data were acquired from 2015 September 24 to 2016 May 30. Here, we simply use the values published in Table 2 of \cite{carmichael19}.

\cite{nowak17} provide 9 RVs measured from spectra acquired with the FIber-fed Echelle Spectrograph (FIES) from 2016 May 15 to 2016 June 29. FIES is located at Roque de los Muchachos Observatory in La Palma, Spain. \cite{nowak17} also publish 3 RV measurements from the Tull Coude Spectrograph, but we omit these from the present work as these data showed significantly more scatter than the FIES or the TRES data (a few hundred \meterps\, from Tull compared to $<50$\,\meterps for TRES and FIES).

\subsection{K2 Data} \label{subsec:k2}
We use photometric time series from the K2 missions as part of our joint modeling of the RM effect. The K2 data are acquired from the Mikulski Archive for Space Telescopes and are processed following the methods outlined in \cite{carmichael19}.

\section{Analysis} \label{sec:analysis}
\begin{figure}[ht!]
\includegraphics[width=0.47\textwidth, trim={0.0cm 0.0cm 0.0cm 0.0cm}]{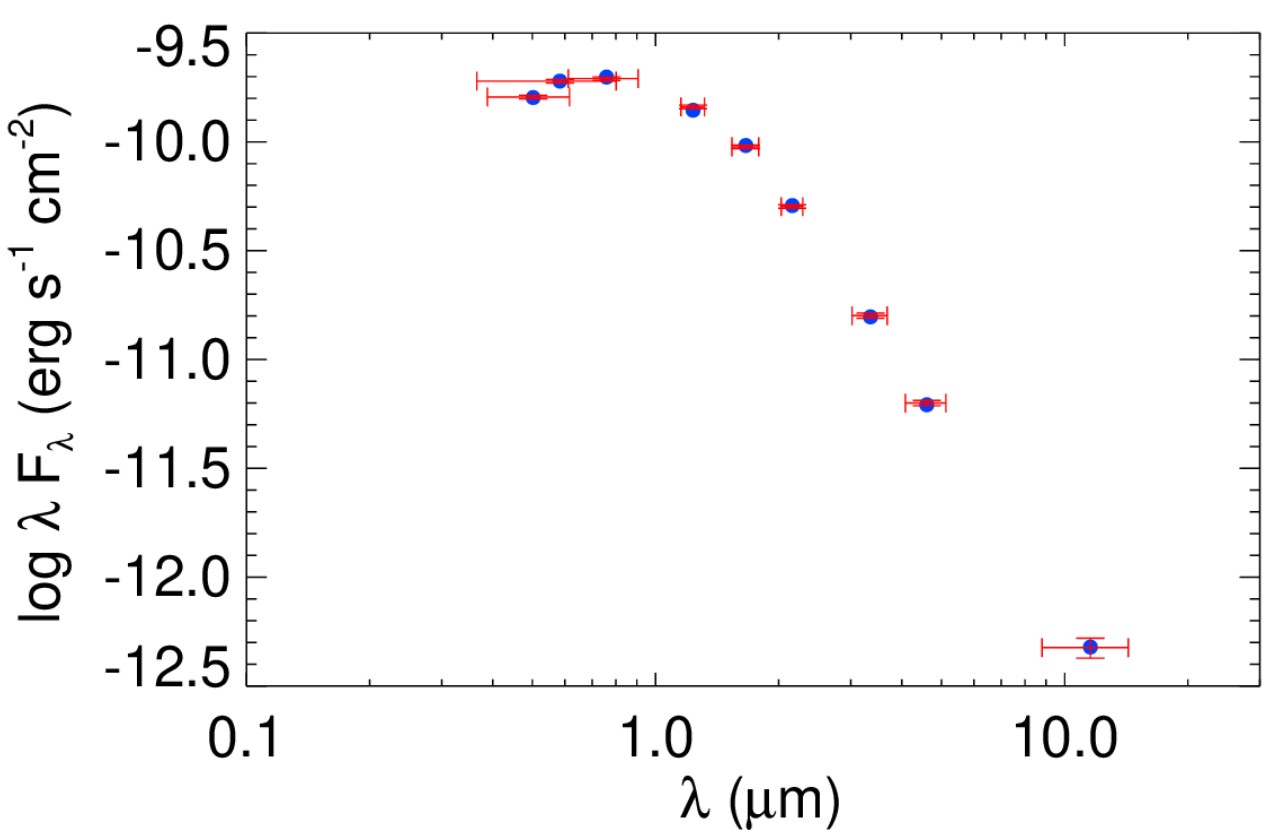}
\caption{\dbf{   SED photometry of CWW 89A. Blue points are the model values from MIST \citep{mist1, mist2, mist3}. Red points are the input data from Gaia \citep{gaia_dr3_magnitude_corr}, 2MASS \citep{2MASS}, and WISE \citep{WISE}.}
\label{fig:SED}}
\end{figure}

\begin{figure}[ht!]
\includegraphics[width=0.47\textwidth, trim={0.0cm 0.0cm 0.5cm 0.0cm}]{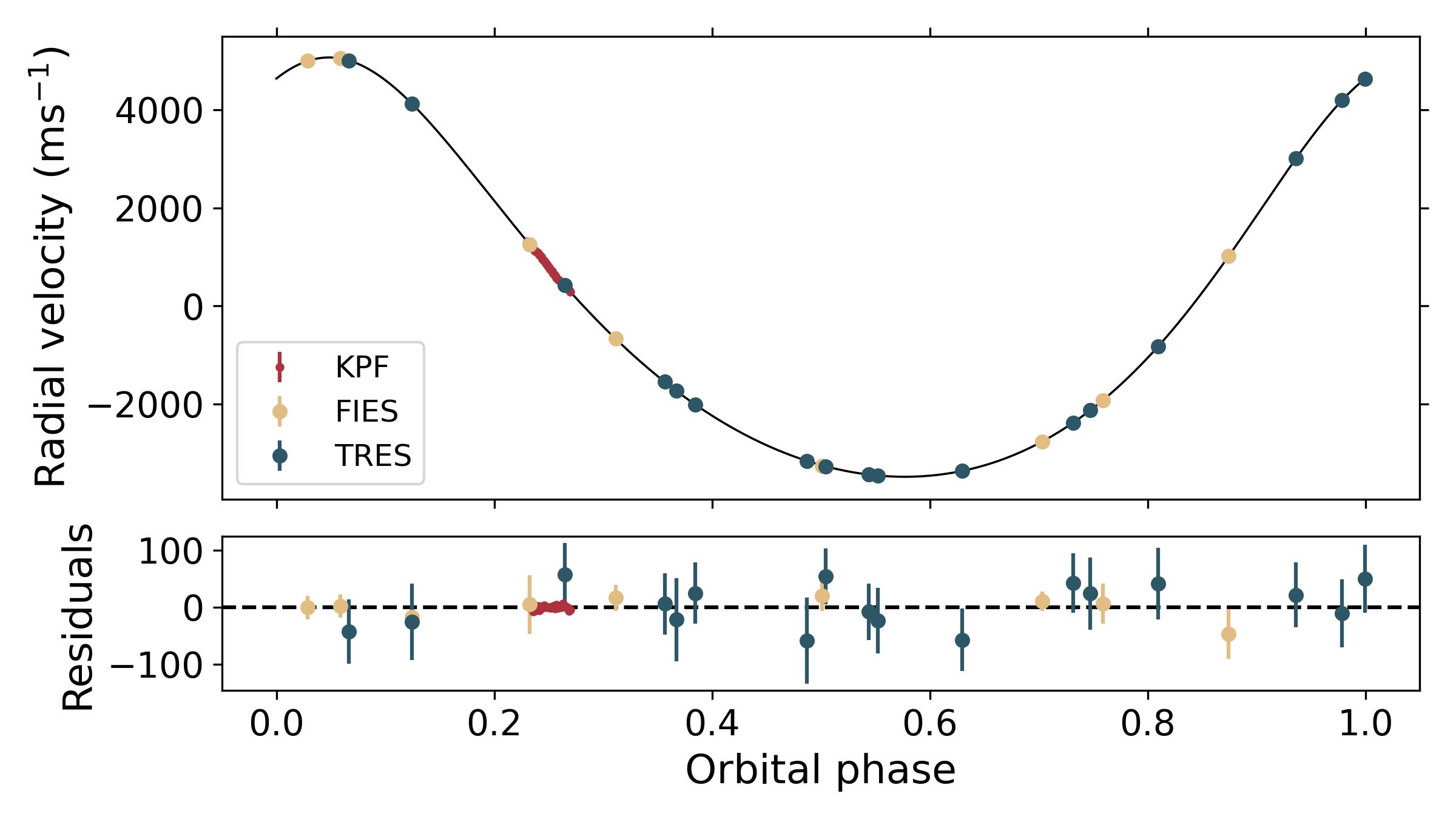}
\includegraphics[width=0.47\textwidth, trim={0.0cm 0.0cm 0.5cm 0.0cm}]{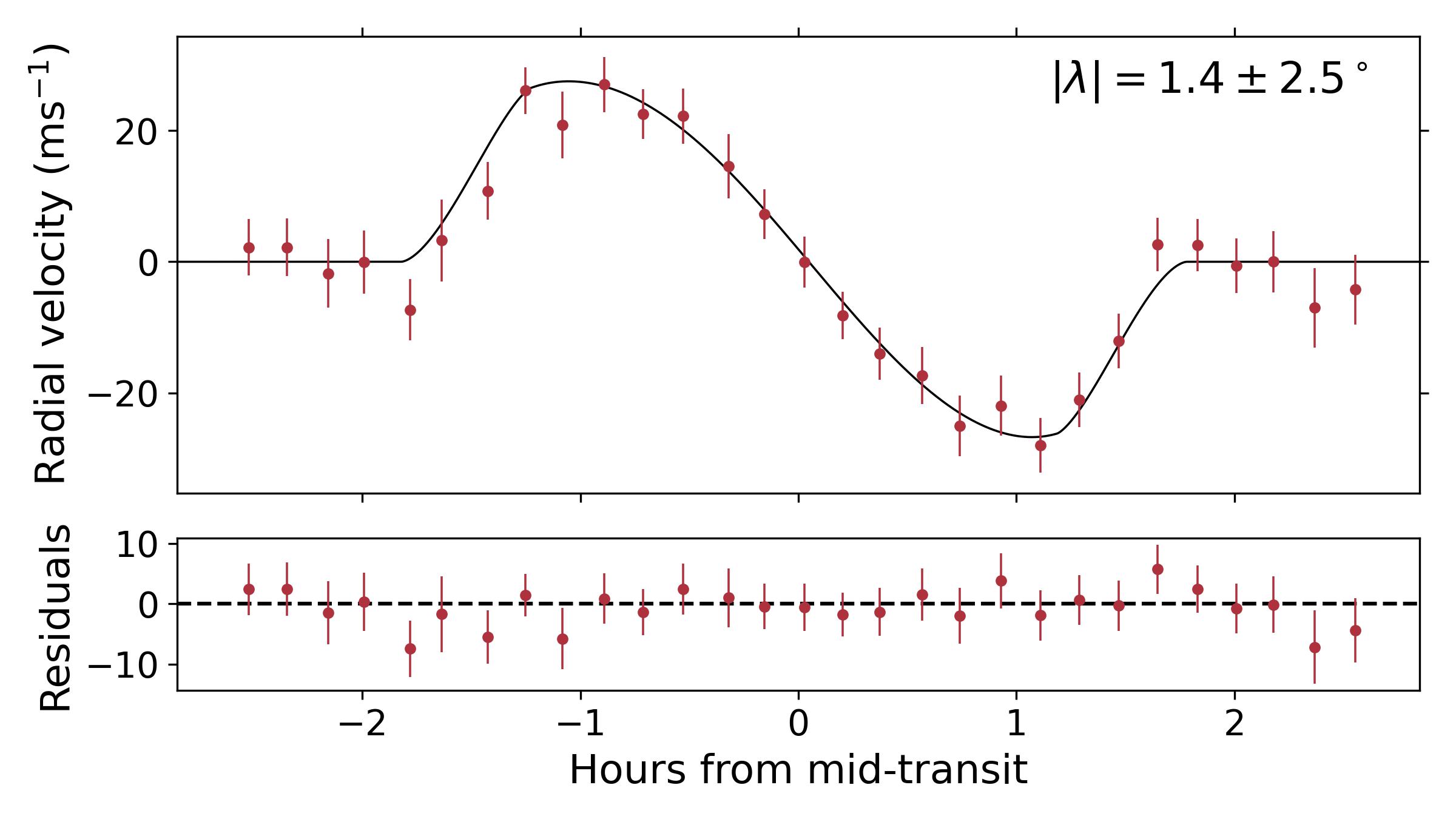}
\includegraphics[width=0.47\textwidth, trim={0.0cm 0.0cm 0.5cm 0.0cm}]{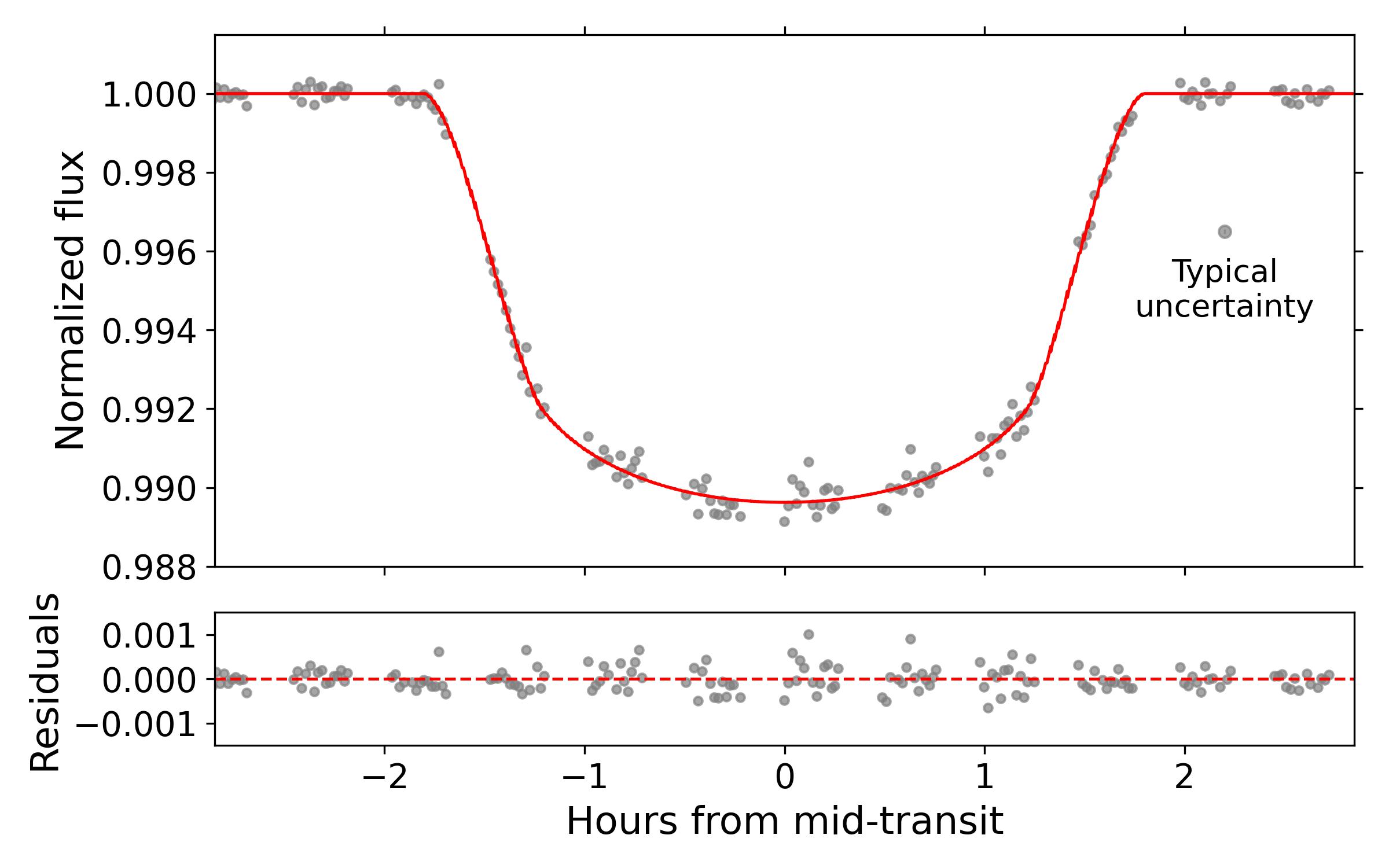}
\caption{\textit{Top:} RV orbital solution for CWW 89A using TRES \citep{carmichael19} and FIES \citep{nowak17} data. \textit{Middle:} KPF in-transit RV data showing the RV anomaly induced by the Rossiter-McLaughlin Effect. \dbf{   The y-axis units in the top and middle panels are in $\rm m\,s^{-1}$. \textit{Bottom:} Phase-fold \textit{K2} transit photometry of CWW 89A with our EXOFASTv2 transit model overlaid.}
\label{fig:rm_plot}}
\end{figure}

We jointly model the KPF, TRES, and FIES RVs, K2 light curves, and host star broadband SED photometry (see Table \ref{tab:sed} using a modified version\footnote{\href{https://github.com/wangxianyu7/EXOFASTv2}{Modified EXOFASTv2}} of EXOFASTv2 \citep{eastman2019} where instead of using a simplified approximation for the RM effect from \cite{ohta2005}, we employ a more robust calculation from \cite{hirano2011}, which accounts for stellar rotation, macroturbulence, thermal broadening, pressure broadening, and instrumental broadening. \dbf{   Figures \ref{fig:SED} \& \ref{fig:rm_plot} show our SED and transit models, respectively. Bandpass magnitudes for the SED fit are shown in Table \ref{tab:sed}. Note again that only the KPF green CCD RV data is included in the RM model. The median uncertainty on the green CCD RVs is 4.3$\rm m\,s^{-1}$ while the median uncertainty on the red CCD RVs is 13.2$\rm m\,s^{-1}$, so we opt to use the better of the two data sets. Using both CCDs results in derived $|\lambda|$ consistent with the green CCD-only fit, but with larger uncertainties.}

\begin{deluxetable}{lccc}
 \tabletypesize{\footnotesize}
 \tablewidth{0pt}
 \tablecaption{\dbf{   Stellar magnitudes used in the SED fitting for CWW 89A. We impose a floor of 0.02 magnitudes on the uncertainty for the Gaia bandpasses to ensure that the SED model is not overconstrained by potentially underestimated photometric errors.} \label{tab:sed}}
 \tablehead{
 \colhead{Name} & \colhead{Description} & \colhead{Value} &  \colhead{Source}}
 \startdata
 $G$\dotfill & Gaia green\dotfill & $12.36 \pm 0.02$  & 1\\
 $G_{BP}$\dotfill & Gaia blue\dotfill & $12.76 \pm 0.02$ & 1\\
 $G_{RP}$\dotfill & Gaia red\dotfill & $11.80 \pm 0.02$  & 1\\
 $J$\dotfill & 2MASS $J$\dotfill & $11.07 \pm 0.02$ & 2\\
 $H$\dotfill & 2MASS $H$\dotfill & $10.73 \pm 0.02$ & 2\\
 $K_S$\dotfill & 2MASS $K_S$\dotfill & $10.67 \pm 0.02$ & 2\\
 WISE1\dotfill & WISE 3.4$\rm \mu m$\dotfill & $10.61 \pm 0.03$  & 3\\
 WISE2\dotfill & WISE 4.6$\rm \mu m$\dotfill & $10.63 \pm 0.03$ & 3\\
 WISE3\dotfill & WISE 12$\rm \mu m$\dotfill & $10.61 \pm 0.12$ & 3\\
 \hline
 \enddata
\tablenotetext{}{References: 1 - \cite{gaia_dr3_magnitude_corr}, 2 - \cite{2MASS}, 3 - \cite{WISE}}
\end{deluxetable}

We set either uniform $\mathcal{U}[a,b]$ or Gaussian $\mathcal{G}[a,b]$ priors on our input parameters. \dbf{   The Gaussian} priors have mean values $a$ with width values $b$ \dbf{   while the uniform priors use $a$ and $b$ and lower and upper bounds, respectively.} We use spectroscopic priors on $T_{\rm eff}$ and [Fe/H] from \cite{carmichael19}, zero-point corrected \citep{Lindegren2018} parallax priors from Gaia DR3, and an upper limit on V-band extinction \citep[$A_V$,][]{av_priors}. \cite{curtis2013} confirm the CWW 89 system to be a probable member of the Ruprecht 147 open cluster, so we additionally set the cluster's age of $2.76 \pm 0.61$ \citep{r147_age} as a prior. We let the RV offsets $\gamma$ for each instrument be free parameters and we include an RV jitter term, $\sigma_j$, for each RV dataset to account for the surface activity of the star. The RV jitter term is based on a white noise model implemented in {\tt EXOFASTv2}. We use values from \cite{carmichael19} for the following parameters to set as starting points for our fit: the orbital period, mid-transit time, orbital inclination, and the companion-to-host radius ratio.

For the RM model, we model the limb-darkening for KPF assuming a $V$ bandpass wavelength range. We fix $v_\alpha$ to 0 (ignore differential rotation) and we use scaling relations for $v_\zeta$ (macroturbulence) from \cite{doyle2014_exofast} and for $v_\xi$ (microturbulence) from \cite{bruntt2010_exofast}. Both $v_\zeta$ and $v_\xi$ have uniform priors. From this joint analysis of the stellar properties, transit photometry, and RVs, we find a projected obliquity angle of $|\lambda|= 1.4\pm2.5^\circ$ for CWW 89Ab. \dbf{   We contextualize this measurement with others for transiting BD systems in Table \ref{tab:bd_obliqs}.}

All stellar parameters we derive are consistent within $<2\sigma$ of those determined in \cite{nowak17} and \cite{carmichael19}. Our full list of derived parameters are shown in Table \ref{tab:cww89a}. We highlight our derived $\teff = 5680\pm86$K for CWW 89A as this is especially relevant in our discussion of the realignment timescale for CWW 89Ab. A summary of other system parameters we derive: $M_\star=1.02\pm0.03\msun$, $R_\star=1.07\pm0.02\rsun$, $\rm [Fe/H]=+0.07\pm0.09$, $M_b=37.3\pm1.6\mj$, $R_b=0.96\pm0.04\rj$, $P_{\rm orb}=5.29$ days. 

\begin{deluxetable*}{ccccc}
\tabletypesize{\footnotesize}
\tablewidth{0pt}
\tablecaption{Table of transiting brown dwarf projected obliquities $\lambda$ and 3D obliquities $\psi$. Values for CWW 89Ab are from this work. \label{tab:bd_obliqs}}
 \tablehead{
 \colhead{Object} & \colhead{Mass ($\mj$)} & \colhead{$|\lambda|$ (degrees)} & \colhead{$\psi$ (degrees)} & \colhead{Source}}
\startdata
HATS-70b & $12.9\pm 1.8$& $8.9\pm 5.6$& - &1\\
GPX-1b & $19.7\pm 1.6$& $6.9\pm 10.0$& - &2\\
CoRoT-3b & $22.3\pm 1.0$ & $37.6^{+10.0}_{-22.3}$ & - &3\\
KELT-1b & $27.3\pm 0.5$& $2.0\pm 16$& - &4\\
CWW 89Ab & $39.2\pm 1.1$& $1.4\pm 2.5$& $15.1^{+15.0}_{-10.9}$ & -\\
WASP-30b & $62.5\pm 1.2$& $7.0^{+19.0}_{-27.0}$& - &5\\
TOI-2119b & $64.4\pm 5.3$& $0.8\pm 1.1$& $15.7\pm 5.6$ &6\\
LP 261-75C & $67.4\pm 2.1$& $4.8^{+11.3}_{-10.2}$& $14.0^{+7.8}_{-6.7}$ &7\\
TOI-2533b & $74.9\pm 5.3$& $7.0\pm 14.0$& - &8\\
\enddata
\tablenotetext{}{References: 1) \cite{hats70b}, 2) \cite{oatmeal1}, 3) \cite{corot3_RM}, 4) \cite{kelt1b}, 5) \cite{wasp30_rm}, 6) \cite{doyle2025}, 7) \cite{brady2025}, 8) \cite{dosSantos2024}}
\end{deluxetable*}

\begin{figure}[ht!]
\includegraphics[width=0.47\textwidth, trim={0.5cm 0.0cm 0.0cm 0.0cm}]{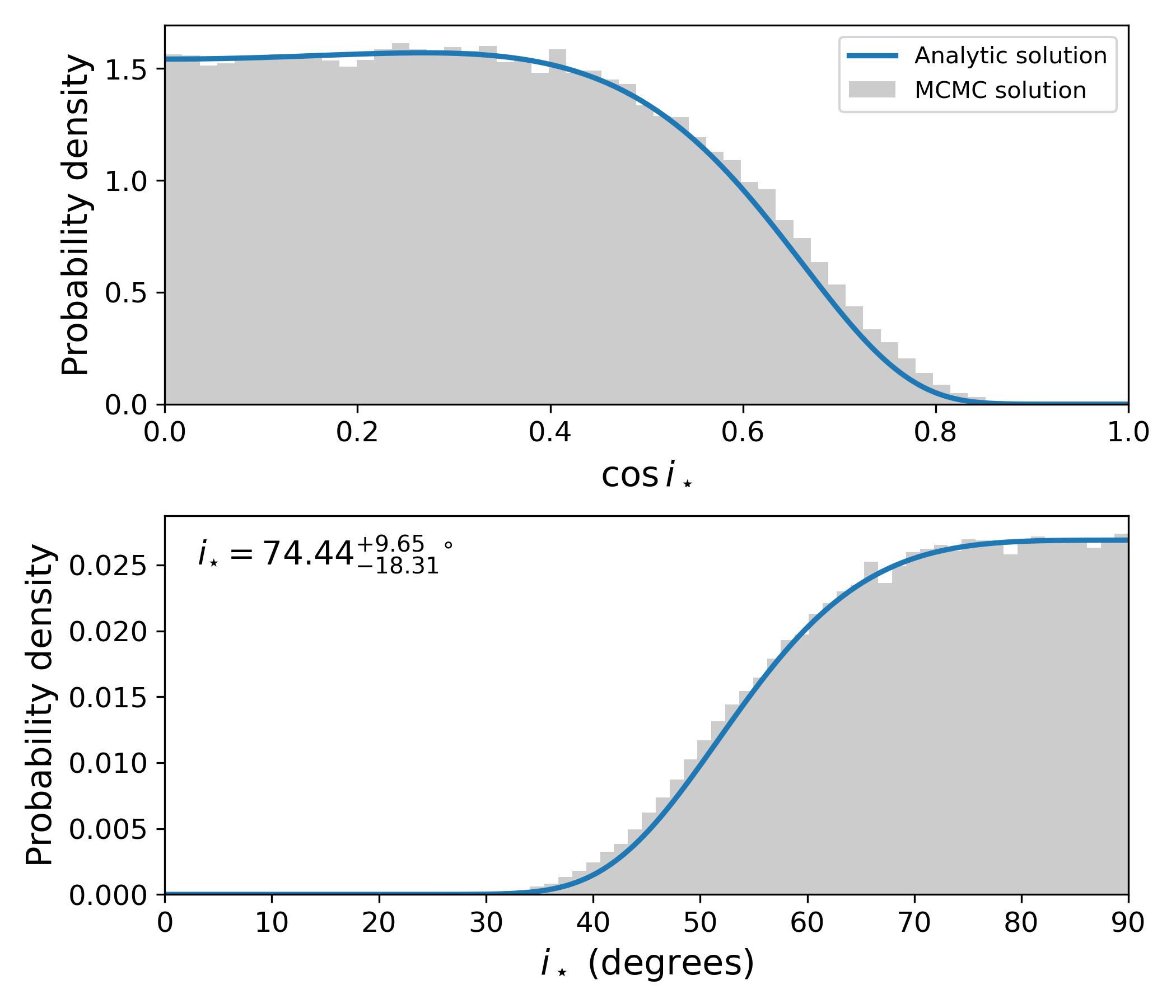}
\caption{The posterior distribution for the stellar inclination angle $i_\star$ for CWW 89A. We use the $P_{\rm rot}$ measured from \cite{nowak17} and the sampling method first presented in \cite{masuda2020} and described in \cite{bowler2023_inclination} to derive an $i_\star=74.4^{+9.7^\circ}_{-18.3}$. We interpret this as aligned with the orbital inclination $i_0$.
\label{fig:stellar_incl}}
\end{figure}

\subsection{The 3D spin-orbit obliquity}
With our newly measured projected obliquity in hand, we may determine the 3D spin-orbit obliquity of the system. This is the angle between the spin axis of the primary star and the orbital axis of the brown dwarf companion. By combining $\lambda$ with the stellar spin axis inclination angle, $i_\star$, and the orbital inclination angle of the BD, $i_0$, the 3D obliquity may be calculated via:

\begin{equation}\label{eq:psi}
\cos{\psi} = \cos{i_\star}\cos{i_0} + \sin{i_\star}\sin{i_0}\cos{\lambda}
\end{equation}

This form of the relationship between $\psi$, $\lambda$, $i_\star$, and $i_0$ follows from \cite{albrecht2022}. The first reported rotation period of CWW 89A comes from \cite{nowak17} on an analysis of the \textit{K2} light curve showing a $P_{\rm rot} = 12.6 \pm 2.1$ days. If we adopt this value and the $v\sin{i_\star} = 4.1\pm 0.4\,(1.0)\, {\rm km\,s^{-1}}$ from \cite{nowak17}, we may derive the stellar inclination as $i_\star=74.4^{+9.7^\circ}_{-18.3}$ (Figure \ref{fig:stellar_incl}). The parenthetical uncertainty on $v\sin{i}$ is the value we adopt given the typical challenges in small $v\sin{i}$ measurements; our RM analysis also provides constraints on $v\sin{i}$ (Table \ref{tab:cww89a}) and both $v\sin{i}$ values are consistent with each other. This derivation follows that first outlined in \cite{masuda2020} and improved on in \cite{bowler2023_inclination} using a bootstrap analysis of the relationship: 
\begin{equation}\label{eq:istar}
    i_\star = \sin^{-1}{\left(\frac{v\sin{i_\star}}{2\pi R_\star/P_{\rm rot}}\right)}
\end{equation}

\begin{figure}[ht!]
\includegraphics[width=0.49\textwidth, trim={1.0cm 0.0cm 0.0cm 0.0cm}]{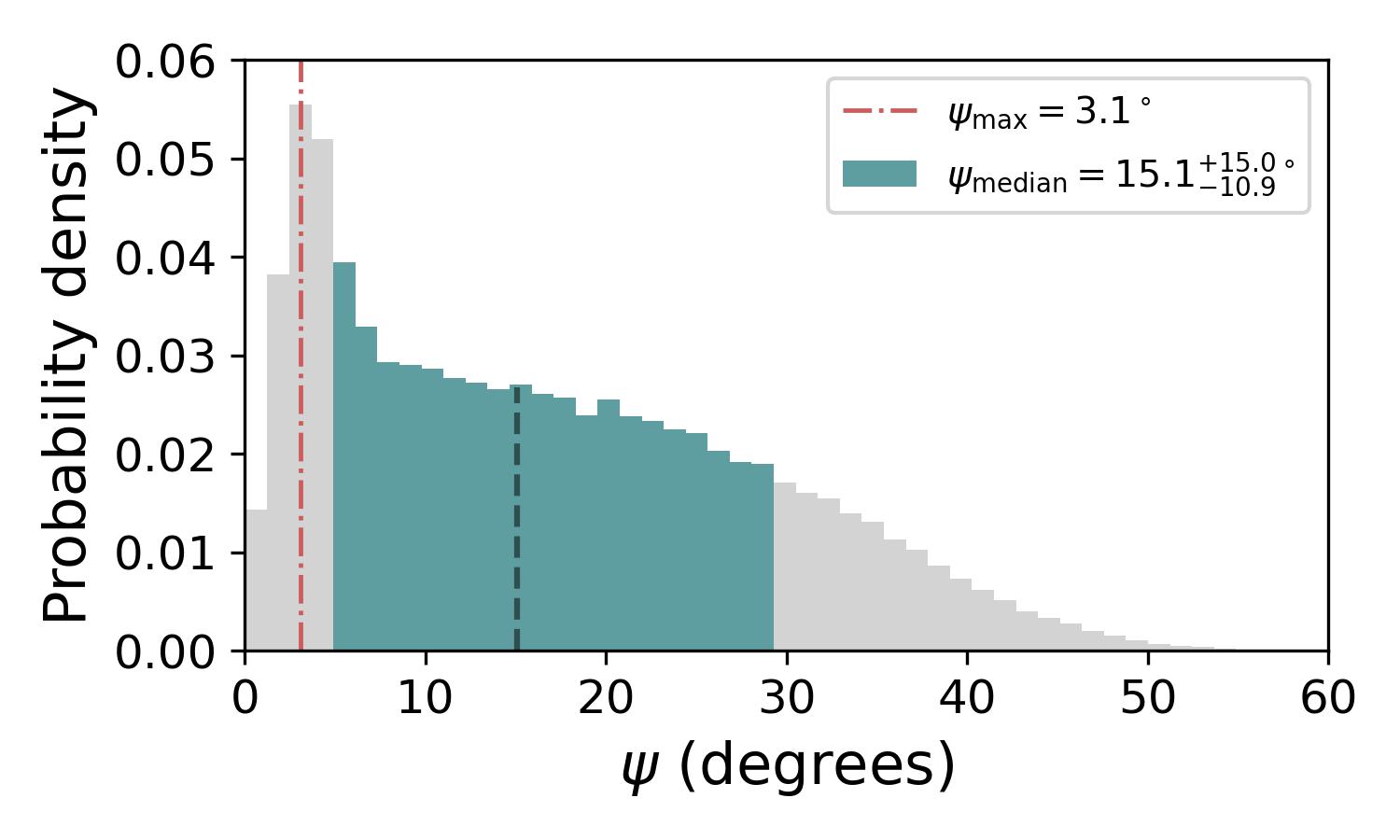}
\caption{The posterior distribution of $\psi$ from sampling in $\cos{\psi}$ parameter space. Given the skewed nature of this distribution \dbf{   (driven by $v\sin{i_\star} \lessapprox v_{\rm rot}$)}, a more conservative estimate of the 3D obliquity is $\psi \leq 30.1^\circ$ with the peak of the distribution at $\psi_{\rm peak} = 3.1^\circ$.
\label{fig:psi_posterior}}
\end{figure}

\dbf{   An artifact of the closeness in value of $v\sin{i_\star}$ and the rotational velocity ($v_{\rm rot} = 2\pi R_\star/P_{\rm rot}$) is a broad distribution in $\cos{i_\star}$; this occurs because the argument of Equation \ref{eq:istar} must be $<1$, or $v\sin{i_\star} < v_{\rm rot}$. This is illustratively explained best in Figure 1, top left panel of \cite{masuda2020}.} In any case, combining the results shown in Figure \ref{fig:stellar_incl} here and Equation \ref{eq:psi} yields a 3D obliquity value of $\psi=15.1^{+15.0^\circ}_{-10.9}$ for CWW 89Ab (Figure \ref{fig:psi_posterior}). This can be compared to $\psi$ measured for other transiting BDs in Table \ref{tab:bd_obliqs}, making CWW 89Ab the third transiting BD with a measurement of $\psi$. We note that our measurements of $\lambda$ and $\psi$ are consistent within 1-$\sigma$ of concurrent measurements taken by \cite{zak2025}.

\section{Discussion} \label{sec:discussion}

\subsection{The CWW 89 system}
A collection of past studies \citep{curtis2013, r147_age, nowak17, cww89a, carmichael19} have helped construct a nearly complete picture for the formation and evolution of the brown dwarf in the CWW 89 system. This section contextualizes the present work and this picture with the aim of further completing it with our new measurements of the spin-orbit obliquity. Summarized here are the key features of the CWW 89 system:
\begin{itemize}
    \item CWW 89A is a $1.0\rm\msun$ star that hosts a distant $0.5\rm\msun$ M-dwarf companion, CWW 89B, with a projected separation of 25 AU, or 81 mas, and $\Delta K = 2.2$ dimmer than CWW 89A \citep{cww89a}
    \item CWW 89 is a probable member of the Ruprecht 147 open cluster \citep{curtis2013} with an age of $2.76\pm 0.61$ Gyr \citep{r147_age}
    \item The brown dwarf, CWW 89b, may have formed via core accretion given evidence presented in \cite{cww89a}; in summary: secondary eclipses from \textit{Spitzer}/IRAC \citep{SpitzerIRAC} indicate an over-luminous atmosphere of CWW 89Ab that is best explained by a temperature inversion caused by a superstellar C/O ratio in the brown dwarf---this C/O ratio favors a core accretion formation scenario
    \item The tidal quality factor of CWW 89b is estimated to have a lower limit of $Q_b \geq 10^{4.15}$ \citep[for details, see section 5 of][]{cww89a}, which yields a circularization timescale $\tau_{\rm circ}> 7.5$ Gyr \citep{carmichael19}
\end{itemize}

\dbf{   Additionally, though we find the conclusions in \cite{cww89a} sound, we note that competing theories on the formation of objects in the brown dwarf mass range tend to favor gravitational instability and fragmentation of the circumstellar disk into proto-brown dwarfs \citep{kratter-lodato2016}. This remains a contention here as no other brown dwarfs near this mass ($\sim$40$\mj$) have evidence for or against a core accretion scenario.} 

\dbf{   Further pulling CWW 89Ab away from a planet-like formation origin is the fact that stellar multiples are often found in hierarchical configurations not dissimilar from the CWW 89 system \citep[e.g.][]{tokovinin2008, duchene2013, moe2017}, which implies that it is more similar to them. A study by \cite{schlaufman2018} also finds evidence that core accretion is challenged to form companions that are more massive than $\gtrsim4\mj$, especially around stars with super-solar metallicity like CWW 89A at $\rm[Fe/H]=+0.2$. These lines of evidence that disfavor a core accretion scenario for CWW 89Ab must be weighed against the findings of \cite{cww89a} for a more complete picture.}

Currently, it is unclear whether or not the secondary star, CWW 89B, is or is not coplanar with CWW 89Ab---an especially important aspect to consider given the implications for the M-dwarf's influence on the orbital evolution of the brown dwarf. \dbf{   We discuss the implications of CWW 89B's non-coplanarity with CWW 89Ab in more detail in Section \ref{sec:secular}.}

\subsection{Evidence for primordial alignment}
\begin{figure*}[ht!]
\centering
\includegraphics[width=0.9\textwidth, trim={1.0cm 0.5cm 1.0cm 0.0cm}]{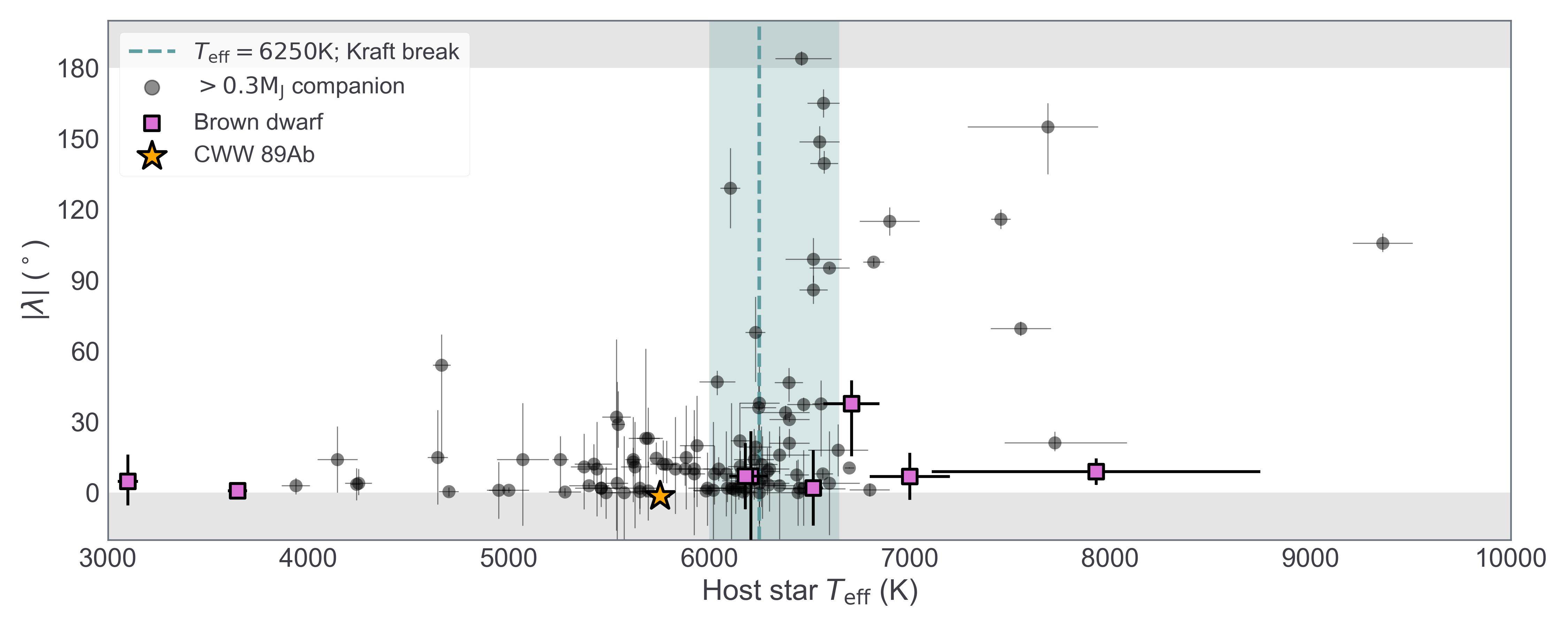}
\includegraphics[width=0.9\textwidth, trim={1.0cm 0.0cm 0.3cm 0.0cm}]{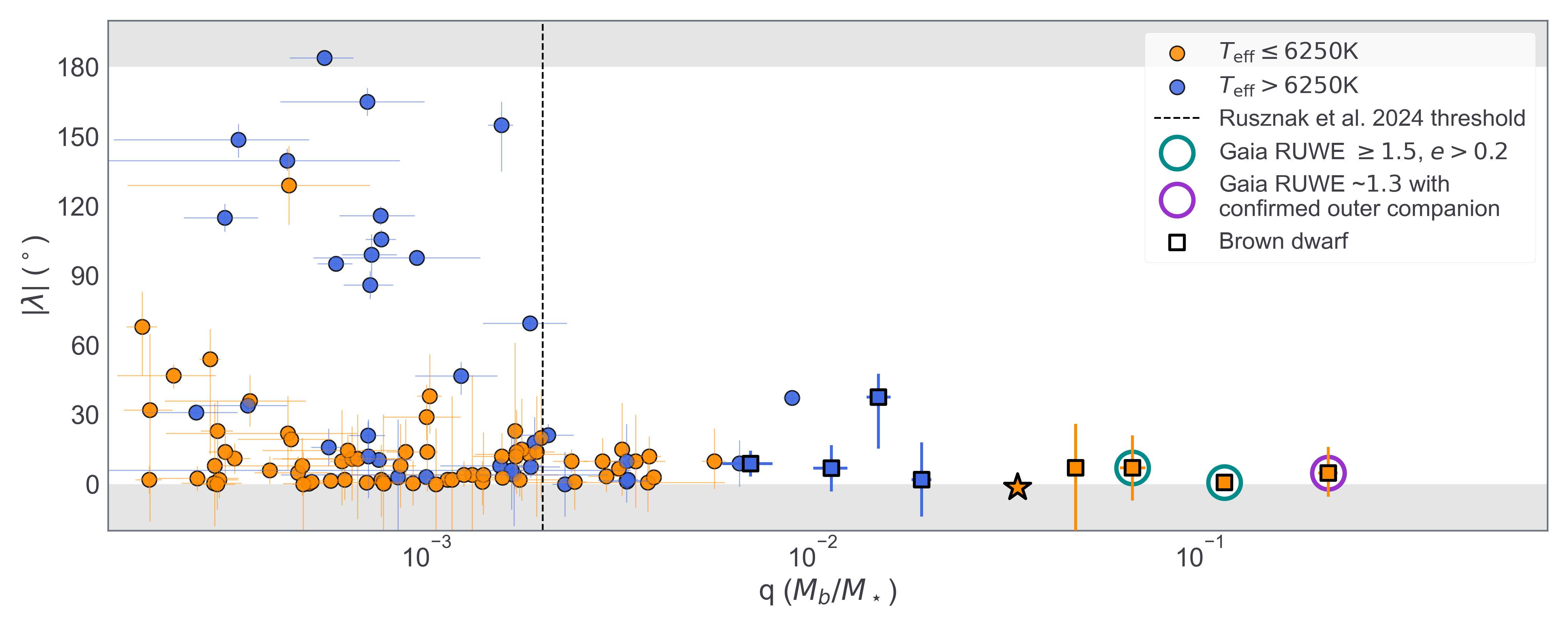}
\caption{\textit{  Top:} $\lambda$ vs $\teff$ for transiting companions $\gtrsim 0.3\mj$. The Kraft Break regime in the vertical shaded region is from \cite{beyer_white2024}, with the traditional 6250K threshold denoted by the dashed line. Brown dwarfs are represented by square symbols. CWW 89Ab is the lowest-mass transiting brown dwarf around a sub-Kraft Break star with its $\lambda$ and $\psi$ measured. \textit{  Bottom:} $\lambda$ vs $q$ ($M_b/M_\star$) for the same sample in the top panel. We highlight the threshold ($q=2\times 10^{-3}$) past which primordial alignment ($\lambda\approx 0^\circ$) dominates the observed distribution \citep[credit to][]{rusznak2024}. Circled points are systems with inner transiting brown dwarfs and confirmed outer companions (CWW 89 and LP 261-75) or Gaia RUWE values above 1.5. \dbf{   Both CWW 89A and LP 261-75 have RUWE values $<1.4$ despite a confirmed stellar companion.} All of the inner brown dwarfs except for the one in LP 261-75 are on eccentric orbits ($e>0.18$), plausibly following a coplanar HEM scenario. The misaligned giant planet system at $q \approx 10^{-2}$ is XO-3b discussed in Section \ref{sec:hier_trips}. Data in both panels are from Figure 2 of \cite{rusznak2024}.
\label{fig:lambda_teff}}
\end{figure*}

Several factors affect the spin-orbit histories and outcomes for giant planet and brown dwarf companions to stars. Three key attributes are: mass ratio $q$, scaled semi-major axis $a/R_\star$, and host star effective temperature. We show Figure \ref{fig:lambda_teff} to illustrate the observed distribution in spin-orbit alignment $\lambda$ as a function of $\teff$ or $q$ (adapted from a similar figure in \cite{rusznak2024}). These factors strongly dictate the circularization timescale $\tau_{\rm circ}$ and realignment timescale $\tau_{\rm CE}$ for giant planets and brown dwarfs, but importantly, these effects appear to be distinct between the two populations. 

From \cite{albrecht2012}, the realignment timescale is:

\begin{equation}\label{eq:tau_ce}
    \tau_{\rm CE} = 10\times q^{-2}\left(\frac{a/R_\star}{40}\right)^6 {\rm Gyr}
\end{equation}

\noindent for convective outer envelope (hence the CE subscript) stars $\teff \lesssim 6250$K. We derive $q=0.035$ and $a/R_\star=12.21$ for CWW 89Ab.


From this, it follows that lower-mass companions (smaller mass ratio $q$) would have longer realignment timescales than higher-mass companions around cool stars. The same conclusion is drawn for companions at larger scaled semi-major axes. In the case of CWW 89A, Equation \ref{eq:tau_ce} indicates the realignment timescale is $\tau_{\rm CE} \approx 10$ Gyr, and greater than the $2.76\pm 0.61$ Gyr age of the system by $>$$3\sigma$. This implies that the spin-orbit alignment of the brown dwarf is primordial. 

However, the provenance of the eccentricity of the brown dwarf is ambiguous given \cite{carmichael19} report the circularization timescale $\tau_{\rm circ}>7.5$ Gyr using the tidal quality factor of $Q_b\gtrsim 10^{4.15}$ for the brown dwarf from \cite{cww89a}. This reported $Q_b$ is based on an extrapolation using rotational angular momentum evolution calculations between a star and giant planet from \cite{leconte2010_tidalQ}. This calculation attempts to account for the exchange in energy via tides that result in the spin-up of the host star rotation rate and the increase in internal heat of the companion. The fact that the circularization timescale is much longer than the system's age while the BD maintains an eccentricity $e=0.19$ means that this is only \textit{possibly} primordial. If the BD was not formed in an eccentric orbit, then its current eccentricity may be explained by interactions with the outer companion, CWW 89B. 

\subsubsection{Coplanar high-eccentricity migration}
This brings us to one theory that could explain the non-zero eccentricity of CWW 89Ab while preserving a zero (aligned) spin-orbit angle. \cite{petrovich2015} presents a framework that plausibly describes the story of CWW 89 via a \textit{coplanar} high-eccentricity migration (coplanar HEM, or CHEM) for the BD driven by the outer M-dwarf. In this scenario, a distant companion could be responsible for guiding the BD inwards to its current configuration without $\lambda$ deviating from zero should the following conditions be satisfied: 

\begin{equation}\label{eq:chem}
    m_{\rm in}/m_{\rm out}(a_{\rm in}/a_{\rm out})^{1/2} \lesssim 0.16
\end{equation}

\noindent where $m$ and $a$ are the respective initial masses and semi-major axes of the inner (CWW 89Ab; $a_{\rm in}=0.06$ AU, $m_{\rm in} = M_b  = 39\mj = 0.04\msun$) and outer companion (CWW 89B; $a_{\rm out}\gtrsim 25$ AU, $m_{\rm out} = M_B = 0.5\msun$). In the case of CWW 89, we are limited with a \textit{projected} separation for CWW 89B, but even in the most conservative scenario (the minimum allowable $a_{\rm out}$) for this framework, the above inequality still holds. \dbf{   This inequality assumes an initial eccentricity of the inner companion $e\gtrsim 0.5$, so this would require the eccentricity of the inner brown dwarf to sufficiently dampen from $e\approx0.5$ to $e=0.2$ for its present day configuration. However, an additional criterion that \cite{petrovich2015} establishes is that if we assume that the initial $e_{\rm in} \approx 0.5$, then for the system to be stable, the initial $e_{\rm out}$ must be similar ($e_{\rm in} \approx e_{\rm out} \approx 0.5$) \textit{and} the mass ratio between the inner and outer companion must be $m_{\rm in}/m_{\rm out} > 0.36$ \citep[see Figure 3 from][]{petrovich2015}. While the present and initially eccentricities of the outer companion are not known, the mass ratio between CWW 89Ab and CWW 89B is known. With an $m_{\rm in}/m_{\rm out} \approx 0.08$, the CWW 89 system falls short of the mass ratio criterion.}

\dbf{   Given this, we consider two alternate scenarios: 1) An initially circular orbit for the BD that was subsequently excited into an eccentric orbit by the outer companion (discussed further in Section \ref{sec:reslock}) or 2) an adherence to the assumption that the BD formed in an eccentric orbit, but in this case, the BD torqued the circumstellar disk due to its high mass. The mass of the BD taken with the evidence that it formed via core accretion \citep{cww89a} may be interpreted as evidence that the circumstellar disk was eccentric and possibly torqued by the BD. Past works have confirmed that such disks have been observed to be eccentric \citep[e.g.][]{kley2006, romanova2023, romanova2024, commercon2024}.}

\dbf{   In both of these scenarios, the spin-orbit obliquity need not be misaligned during the inward high-eccentricity migration process of the BD. An example of this may be seen in the well-aligned ($\lambda=1.2\pm 2.8^\circ$) but eccentric ($e=0.72$) gas giant TOI-3362b \citep{espinoza2023_CHEM}. When taking these various factors into consideration, we are unable to decisively conclude whether or not CWW 89 is stable under the CHEM framework despite its cursory appearance of being an exemplary CHEM system like TOI-3362. Even so, the extreme mass and separation scales of CWW 89  \citep[compared to those considered in][]{petrovich2015} are not yet thoroughly tested in the CHEM framework and may not be completely ruled out.}

\subsubsection{Other massive hierarchical triple systems}\label{sec:hier_trips}
There are 4 other transiting systems that feature an inner massive ($\gtrsim 10\mj$) companion with a measured spin-orbit angle and with confirmed or plausible outer stellar companions. These are TOI-2119 \citep{canas2021, carmichael2022, doyle2025}, and TOI-2533 \citep{Schmidt2023, dosSantos2024}, LP 261-75 \citep{irwin18, brady2025}, and XO-3 \citep{xo3b, rusznak2024}. Given how suitably the coplanar high-eccentricity migration scenario could explain the CWW 89 system, we examine how it might apply to these other systems.

\begin{itemize}
    \item \textbf{Aligned inner brown dwarf companions:} Findings from both \cite{doyle2025} and \cite{dosSantos2024} for TOI-2119b and TOI-2533b, respectively, tell a closely-related story to that of the CWW 89 system. These systems share certain similar properties with CWW 89Ab with some distinctions that we consider here. Both TOI-2119b and TOI-2533b are massive (60-75$\mj$; unlike CWW 89Ab at 39$\mj$) BDs in eccentric orbits ($e=$0.25-0.34; similar to CWW 89Ab at $e=$0.19) between 6.6 and 7.3 days (longer than  CWW 89Ab at 5.5 days). As shown in Table \ref{tab:bd_obliqs}, TOI-2119b and TOI-2533b are spin-orbit aligned like CWW 89Ab. 
    
    What is absent from our current knowledge of these other 2 BDs is a confirmation of an outer stellar companion like with the detection of CWW 89B. \cite{doyle2025} cite the relatively large re-normalized unweighted error \citep[RUWE,][]{ruweDR2, ruwedr3} from the Gaia mission of 1.93 for TOI-2119 as an indicator of an unresolved massive companion. An RUWE $>$ 1.4 indicates the strong potential for a binary system \citep{ruweDR2, Krolikowski2021_RUWE}. 
    
    Notably, TOI-2119 is an M-dwarf and could join other M-dwarf primary hierarchical triple systems with an inner transiting BD and outer secondary M-dwarf if the high RUWE is \dbf{   strong evidence} of an outer companion. These other hierarchical systems are: LHS 6343 \citep{johnson11_bd}, NLTT 41135 \citep{irwin10}, NGTS-7 \citep{jackman2019}, NGTS-28 \citep{ngts28A}, LP 261-75 \citep{irwin18, brady2025}, \dbf{   and TOI-5389 \citep{larsen2025}. These 6 systems are a part of the total 14 M-dwarf primaries known to host transiting BDs ($\sim43\%$); the other 8}, TOI-2119 included, do not have confirmed outer companions. This emerging trend appears to show BDs that transit M-dwarfs to have a reasonable chance to be part of a hierarchical system with an additional outer M-dwarf. We note that one recent study by \cite{cifuentes2025} indicates that the M-dwarf multiplicity fraction for separations of 0.01 to $10^5$ AU ranges from 26\% to 42\% depending on the confirmation of candidate stellar companions. However, we do not robustly confirm that this is a trend for M-dwarf hosts to transiting BDs here, so the best quantitative evidence for TOI-2119 to host an unresolved M-dwarf companion is its inflated RUWE. 
    
    The TOI-2533 system has an RUWE of 1.57, but is an F8 star. With only \dbf{   circumstantial} evidence for the existence of an outer companion in the TOI-2119 and TOI-2533 systems via the Gaia RUWE exceeding a value of 1.4, it is not clear if an outer companion would extend the shared characteristics between TOI-2119, TOI-2533, and CWW 89, thus fitting the same coplanar HEM scenario from \cite{petrovich2015}. \dbf{   We point out that the RUWE for CWW 89A is only 1.2.} 
    
    \dbf{   A study by \cite{castro-ginard2024_RUWE} demonstrated that the nominal RUWE binarity threshold of 1.4 can vary from $\sim$1.15 to $\sim$1.37 depending on the location in the sky (e.g. for crowded field calibrations) of the system due to the scanning angle of Gaia. There are also other proxies from Gaia for binarity are used in tandem with RUWE \citep[e.g the astrometric acceleration;][]{brandt2021, JZhang2024}. Given this, we defer to a clear direct detection when arguing for the existence of an outer companion in this context.}
    
    Lastly, if TOI-2119 and TOI-2533 host a secondary M-dwarf companion, we could conclude that the condition in Equation \ref{eq:chem} would be satisfied, since we could assume the hypothetical companion orbits exterior to the brown dwarf. \dbf{   Given the much higher masses of TOI-2119b and TOI-2533b compared to CWW 89Ab, the mass ratio criterion from \cite{petrovich2015} would be easier to satisfy if the potential unseen companions in these 2 other systems were roughly $0.2\msun$.}

    \item \textbf{Misaligned inner super-Jupiter companion:} XO-3b is a massive $12\mj$ planet in a \textit{misaligned} ($\lambda=37.8\pm3.7.^\circ$) and eccentric ($e=0.28$) 3-day orbit \citep{xo3b}. This makes XO-3 the singular exception to the observed trend in aligned systems with high mass ratio (Figure \ref{fig:lambda_teff}). \cite{rusznak2024} are confirm the presence of a stellar companion via high-resolution imaging at 90AU for this system and highlight the RUWE of 1.25. \cite{rusznak2024} argue that this outer stellar companion may have been the dominant factor in the misalignment of XO-3b over the effective temperature of the primary star being above the Kraft Break. This system is the only outlier among other massive companions ($M_b \gtrsim 10\mj$) with well-determined $\lambda$ and it is unclear if the outer companion plays a role in this in the same fashion the outer companion to CWW 89A does. 
\end{itemize}

\cite{rusznak2024} argue that the increase in planet-to-star mass ratio correlates with a preference for aligned spin-orbit angles for \textit{single-star} systems. Their conclusions are intended to be valid for single-star systems because these are the objects that they limited their sample to. They provide an explanation for this spin-orbit alignment phenomenon in the circumstellar disks from which these giant planets and brown dwarfs grew. For those systems that experience dynamically disruptive events early in their formation process, especially in low-mass disks with less mass available to accrete, they may experience a kick that can either or both drive an increase in eccentricity (i.e. high-eccentricity migration) and a misalignment with the plane of the disk and host star, thus depriving the planet of any additional material to grow in mass.

This argument is presented to explain the ubiquity of misaligned sub-Saturn planets around stars above and below the Kraft Break. The opposite of this is seen for brown dwarf-mass and even super-Jupiter mass (i.e. $\gtrsim 2\mj$) companions where every system is aligned around hot and cool stars (Figure \ref{fig:lambda_teff}). It is then the Jupiter-mass planets which represent a transitional region in mass where the effects of early-time dynamical excitation affects these planets less strongly than sub-Saturns, but more strongly than super-Jupiters and brown dwarfs. This is a summary of the argument established in \cite{rusznak2024} and it is clear that the transiting brown dwarf population follows this very closely; CWW 89Ab being no exception save for the fact that it is not a single-star system as \cite{rusznak2024} require for their argument to be valid (allowing them to exclude the XO-3 system).

Of the super-Jupiter and brown dwarf systems with measured spin-orbit angles and RUWE $\gtrsim 1.25$, only 1 (XO-3b) in 4 total (CWW 89, TOI-2119, and TOI-2533) is misaligned. As this particular sample continues to grow, we expect it to reveal how the presence of outer companions to transiting brown dwarfs affect spin-orbit alignment. This work and the recent few confirming the aligned orbits of transiting brown dwarfs with the host star's spin axis \dbf{   may have growing implications at an underlying obliquity distribution for these massive companions is consistent with $\lambda \approx 0^\circ$}. This would follow the conclusions that \cite{hebrard2011} establish early in the era of spin-orbit measurements for giant planet and subsequent works by \cite{hixenbaugh2023, gan2024, oatmeal1} would later reinforce. 

\dbf{   Though the total sample for transiting brown dwarfs with measurement spin-orbit obliquities is quite small at 9 systems, the sample of transiting companions $\gtrsim 2\mj$ is larger by a factor of 3. For this sample, the evidence is stronger for a $\lambda\approx0^\circ$ spin-orbit obliquity and future investigations may find this parameter space of $\lambda$ vs. $q$ appropriate for relating the super-Jupiter and brown dwarf populations to one another.}

\subsection{Evidence against primordial misalignment}\label{sec:reslock}
\subsubsection{\dbf{   Resonance locking}}
A separate mechanism that describes the realignment of planetary companions is known as ``resonance locking'' \citep{witte1999, savonije2008}. This is when the orbital motion of the companion couples to the host star's gravity mode (\textit{g}-mode) oscillation frequency, resulting in a gradual relative alignment of the companion's orbit and star's spin axis. \cite{zanazzi2024} simulate a resonance lock scenario using a sunlike star and a $1\mj$ companion to explore different timescales over which the companion's spin-orbit obliquity would realign from misaligned initial values of $\psi$ from $45^\circ$ to $90^\circ$. One important relationship \cite{zanazzi2024} show is between the realignment timescale and scaled semi-major axis $a/R_\star$, where resonance lock is not possible between a 2.5 Gyr old, $1\msun$ star (like CWW 89A) and a $1\mj$ companion at $a/R_\star>10$ within 10 Gyr (Figure 15 in \cite{zanazzi2024}). This scenario is for an initial $\psi=90^\circ$. 

In our case, the companion is almost a factor of 40 more massive, which means that the brown dwarf has an increased tidal potential energy and thus is able to act on the star at larger $a/R_\star$ than a Jupiter-mass companion. However, even when considering the $a/R_\star\gtrsim 12$ for CWW 89Ab, the shortest plausible timescale for realignment from $\psi=90^\circ$ (when extrapolating from \cite{zanazzi2024}) is roughly 9-10 Gyr for CWW 89A, significantly longer than the 2.5 Gyr age of CWW 89. We also highlight the fact that the brown dwarf did not form in its current close orbit, meaning that the coupling of its orbital motion to the \textit{g}-mode oscillations were likely even weaker for a portion of the system's main sequence lifetime. Given this, we find it unlikely that CWW 89A was realigned with the orbital of CWW 89Ab via resonance locking\dbf{   , but we acknowledge that, like with the CHEM framework, the limits of resonance locking has not been tested to these extremes in companion mass}.

\subsubsection{\dbf{   Secular perturbations from CWW 89B}}\label{sec:secular}
\dbf{   The nature of CWW 89B's orbital inclination has profound impacts on the potential alignment of the brown dwarf. A misaligned outer companion would likely make the BD susceptible to secular perturbations like the von Zeipel-Kozai-Lidov effect \citep{zeipel1910, kozai1962, lidov1962} where the spin-orbit alignment $\lambda$ of the BD would oscillate between misaligned and aligned configurations driven by CWW 89B \citep{naoz2016}. This would imply that the our measurement of $\lambda$ took place during a part in this cycle when the BD's spin-orbit angle is near-zero, leading us to misinterpret the system as primordially and perpetually aligned. Whether or not these oscillations excite only the BD's eccentricity or both the eccentricity \textit{and} spin-orbit angle is unclear without knowing CWW 89B's orbital inclination.}

\section{Summary} \label{sec:summary}
We report the aligned orbit of the transiting brown dwarf, CWW 89Ab, from our analysis of KPF in-transit RV data measuring the Rossiter-McLaughlin effect. Our main findings are:

\begin{itemize}
    \item CWW 89Ab has a projected spin-orbit alignment of $|\lambda|=1.4\pm2.5^\circ$.
    \item Given the stellar rotation period and projected stellar rotational velocity from \cite{nowak17}, we find CWW 89Ab has a 3D spin-orbit obliquity of $\psi = 15.1^{+15.0^\circ}_{-10.1}$ with the posterior distribution peaking at $\psi_{\rm peak}=3.1^\circ$.
    \item We confirm trends described in \cite{oatmeal1} and \cite{rusznak2024} that all transiting brown dwarfs in single- or multi-star systems with measured spin-orbit obliquities are consistent with $\lambda=0^\circ$ (aligned) regardless of the primary $\teff$.
    \item Including CWW 89Ab, total number of transiting BDs with measured $\lambda$ is 9 and the number with $\psi$ measured is 3.
\end{itemize}

We find the trend of massive aligned companions to continue with CWW 89A, plausibly following the orbital evolutionary mechanisms outlined in \cite{petrovich2015} and \cite{rusznak2024}, where a high circumstellar disk mass may have enabled the growth of the brown dwarf into a mass regime where it is resistant to dynamical kicks perpendicular to the disk plane that would have misaligned the brown dwarf's orbit were it closer to or less than the mass of Jupiter. What would then follow is a coplanar high-eccentricity migration induced by the outer companion, assuming the outer companion is coplanar with the brown dwarf. Given the long realignment timescale $\tau_{\rm CE} \approx 10$ Gyr between the brown dwarf and primary star and the long circularization timescale $\tau_{\rm circ} \gtrsim 7.5$ Gyr relative to the system's age of 2.8$\pm$0.6 Gyr, we may conclude that CWW 89Ab formed in an eccentric, aligned orbit \dbf{   if CWW 89B is coplanar with CWW 89Ab. If this is not the case, then other mechanisms like von Zeipel-Kozai-Lidov oscillations would be favored over coplanar high-eccentricity migration.} 

Following the conclusion from \cite{cww89a}, if we also assume CWW 89Ab formed via core accretion, \dbf{   which is likely challenging at this mass}, then a more complete picture of CWW 89Ab's formation is: core accretion-driven formation and growth in an aligned, eccentric orbit within the circumstellar disk, followed by a coplanar high-eccentricity migration via its outer M-dwarf companion, CWW 89B, inward to its current architectural configuration. 

One key missing piece of information is \dbf{   orbital inclination} of CWW 89B \dbf{   with respect to} CWW 89A. Future observations with facilities like Keck II/NIRC2 could provide constraints on the orbit CWW 89B \dbf{   to determine this angle}. Additional high-resolution imaging could also be performed on TOI-2119 and TOI-2533 to search for outer companions and the source of the high RUWE value ($\gtrsim1.6$) for both of these stars.

\begin{acknowledgments}
The authors wish to recognize and acknowledge the very significant cultural role and reverence that the summit of Maunakea has always had within the Native Hawaiian community. We are most fortunate to have the opportunity to conduct observations from this mountain.

Some of the data presented herein were obtained at Keck Observatory, which is a private 501(c)3 non-profit organization operated as a scientific partnership among the California Institute of Technology, the University of California, and the National Aeronautics and Space Administration. The Observatory was made possible by the generous financial support of the W. M. Keck Foundation.

This research was carried out, in part, at the Jet Propulsion Laboratory and the California Institute of Technology under a contract with the National Aeronautics and Space Administration and funded through the President’s and Director’s Research \& Development Fund Program.

This research has made use of the NASA Exoplanet Archive, which is operated by the California Institute of Technology, under contract with the National Aeronautics and Space Administration under the Exoplanet Exploration Program.

The research was carried out, in part, at the Jet Propulsion Laboratory, California Institute of Technology, under a contract with the National Aeronautics and Space Administration (80NM0018D0004) and funded through the President’s and Director’s  Research \& Development Fund Program. 

TWC is supported by an NSF MPS-Ascend Postdoctoral Fellowship under award 2316566.
D.H. acknowledges support from the Alfred P. Sloan Foundation and the National Aeronautics and Space Administration (80NSSC22K0781).

\end{acknowledgments}

\vspace{5mm}




\bibliography{citations, obliquity, transitingBDs}{}
\bibliographystyle{aasjournal}

\startlongtable
\begin{deluxetable*}{lccccccc}
\tablecaption{MIST median values and 68\% confidence interval for CWW 89A. Here, $\mathcal{U}$[a,b] is the uniform prior bounded between $a$ and $b$, and $\mathcal{G}[a,b]$ is a Gaussian prior of mean $a$ and width $b$.}
\tablehead{\colhead{~~~Parameter}  & \colhead{Description} & \colhead{Priors} & \multicolumn{5}{c}{Values}}

\startdata
\smallskip\\\multicolumn{2}{l}{Stellar Parameters:}&\smallskip\\
~~~~$M_*$\dotfill &Mass (\msun)\dotfill & - & $1.018^{+0.027}_{-0.026}$\\
~~~~$R_*$\dotfill &Radius (\rsun)\dotfill & - &$1.067\pm0.022$\\
~~~~$L_*$\dotfill &Luminosity (\lsun)\dotfill & - &$0.973^{+0.034}_{-0.037}$\\
~~~~$\rho_*$\dotfill &Density (cgs)\dotfill & - &$1.186^{+0.024}_{-0.022}$\\
~~~~$\log{g}$\dotfill &Surface gravity (cgs)\dotfill & - &$4.391\pm0.017$\\
~~~~$T_{\rm eff}$\dotfill &Effective temperature (K)\dotfill & $\mathcal{G}$[5699,200] &$5553^{+82}_{-86}$\\
~~~~$[{\rm Fe/H}]$\dotfill &Metallicity (dex)\dotfill & $\mathcal{G}$[0.12,0.1] &$0.067\pm0.090$\\
~~~~$Age$\dotfill &Age (Gyr)\dotfill &$\mathcal{G}$[2.76,0.61] &$2.52\pm0.61$\\
~~~~$A_V$\dotfill &V-band extinction (mag)\dotfill & $\mathcal{U}$[0,0.278] &$0.262^{+0.023}_{-0.045}$\\
~~~~$\sigma_{SED}$\dotfill &SED photometry error scaling \dotfill & - &$1.13^{+0.47}_{-0.29}$\\
~~~~$\varpi$\dotfill &Parallax (mas)\dotfill & $\mathcal{G}$[3.286,0.018] &$3.282\pm0.018$\\
~~~~$d$\dotfill &Distance (pc)\dotfill & - &$304.6\pm1.7$\\
~~~~$V_{\rm beta}$\dotfill &Gaussian dispersion (m/s)\dotfill & $\mathcal{G}$[4000,1000] &$4080^{+1000}_{-990}$\\
~~~~$V_{\rm gamma}$\dotfill &Lorentzian dispersion or Differential rotation (m/s)\dotfill & $\mathcal{U}$[1000,1000] (fixed) &$1000.00$\\
~~~~$V_{\rm zeta}$\dotfill &Macroturbulence dispersion (m/s)\dotfill & - &$3050\pm930$\\
~~~~$V_{\rm xi}$\dotfill &Microturbulence dispersion (m/s)\dotfill & - &$1030^{+880}_{-660}$\\
\smallskip\\\multicolumn{2}{l}{Brown dwarf Parameters:}&&b\smallskip\\
~~~~$P$\dotfill &Period (days)\dotfill & - &$5.2926298^{+0.0000041}_{-0.000086}$\\
~~~~$R_P$\dotfill &Radius (\rj)\dotfill & - &$0.963^{+0.037}_{-0.033}$\\
~~~~$M_P$\dotfill &Mass (\mj)\dotfill & - &$37.25^{+1.6}_{-0.98}$\\
~~~~$T_C$\dotfill &Model Time of conjunction$^{1,2}$ (\tjdtdb)\dotfill & - &$2457341.037551^{+0.000095}_{-0.000094}$\\
~~~~$T_0$\dotfill &Obs time of min proj sep$^{3,4,5}$ (\bjdtdb)\dotfill & - &$2457346.329698^{+0.000094}_{-0.00011}$\\
~~~~$a$\dotfill &Semi-major axis (AU)\dotfill & - &$0.06048^{+0.0013}_{-0.00079}$\\
~~~~$i_0$\dotfill &Inclination (Degrees)\dotfill & - &$88.01^{+0.14}_{-0.13}$\\
~~~~$b$\dotfill &Transit impact parameter \dotfill & - &$0.429^{+0.013}_{-0.015}$\\
~~~~$e$\dotfill &Eccentricity \dotfill & - &$0.1892^{+0.0024}_{-0.0029}$\\
~~~~$e\cos{\omega_*}$\dotfill & \dotfill & - &$0.1836^{+0.0023}_{-0.0026}$\\
~~~~$e\sin{\omega_*}$\dotfill & \dotfill & - &$-0.0450\pm0.0032$\\
~~~~$\omega_*$\dotfill &Arg of periastron (Degrees)\dotfill & - &$-13.77^{+0.93}_{-0.89}$\\
~~~~$|\lambda|$\dotfill &Projected Spin-orbit alignment (Degrees)\dotfill & - &$1.4^{+2.1}_{-2.5}$\\
~~~~$v\sin{i_\star}$\dotfill &Projected rotational velocity (m/s)\dotfill & - &$4690^{+510}_{-350}$\\
~~~~$T_{\rm eq}$\dotfill &Equilibrium temp$^{6}$ (K)\dotfill & - &$1268\pm15$\\
~~~~$K$\dotfill &RV semi-amplitude (m/s)\dotfill & - &$4272^{+18}_{-19}$\\
~~~~$R_P/R_*$\dotfill &Radius of planet in stellar radii \dotfill & - &$0.09283^{+0.00051}_{-0.00057}$\\
~~~~$a/R_*$\dotfill &Semi-major axis in stellar radii \dotfill & - &$12.21^{+0.16}_{-0.14}$\\
~~~~$\delta$\dotfill &$\left(R_P/R_*\right)^2$ \dotfill & - &$0.008617^{+0.00010}_{-0.00011}$\\
~~~~$\delta_{\rm Kepler}$\dotfill &Transit depth in Kepler (frac)\dotfill & - &$0.01076^{+0.00011}_{-0.000100}$\\
~~~~$\tau$\dotfill &In/egress transit duration (days)\dotfill & - &$0.02409^{+0.00090}_{-0.00095}$\\
~~~~$T_{14}$\dotfill &Total transit duration (days)\dotfill & - &$0.15017^{+0.00052}_{-0.00050}$\\
~~~~$T_{FWHM}$\dotfill &FWHM transit duration (days)\dotfill & - &$0.12609^{+0.00067}_{-0.00063}$\\
~~~~$\rho_P$\dotfill &Density (cgs)\dotfill & - &$23.4^{+1.6}_{-1.4}$\\
~~~~$logg_P$\dotfill &Surface gravity (cgs)\dotfill & - &$4.764^{+0.019}_{-0.016}$\\
~~~~$T_P$\dotfill &Time of Periastron (\tjdtdb)\dotfill & - &$2457339.830^{+0.014}_{-0.013}$\\
~~~~$T_A$\dotfill &Time of asc node (\tjdtdb)\dotfill & - &$2457339.9660\pm0.0050$\\
~~~~$T_D$\dotfill &Time of desc node (\tjdtdb)\dotfill & - &$2457342.7668^{+0.0087}_{-0.0097}$\\
~~~~$M_P\sin i_0$\dotfill &Minimum mass (\mj)\dotfill & - &$37.23^{+1.6}_{-0.99}$\\
~~~~$M_P/M_*$\dotfill &Mass ratio \dotfill & - &$0.03492^{+0.00053}_{-0.00076}$\\
\hline
\smallskip\\\multicolumn{2}{l}{Wavelength Parameters:}&Kepler&V\smallskip\\
~~~~$u_{1}$\dotfill &Linear limb-darkening coeff \dotfill &$0.433^{+0.035}_{-0.032}$&$0.502^{+0.052}_{-0.057}$\\
~~~~$u_{2}$\dotfill &Quadratic limb-darkening coeff \dotfill &$0.253\pm0.047$&$0.253^{+0.054}_{-0.050}$\\
\smallskip\\\multicolumn{2}{l}{Telescope Parameters:}&FIES&KPF&TRES\smallskip\\
~~~~$\gamma_{\rm rel}$\dotfill &Relative RV Offset (m/s)\dotfill &$45642.2^{+9.6}_{-9.0}$&$46716^{+18}_{-230}$&$3481^{+12}_{-13}$\\
~~~~$\sigma_J$\dotfill &RV Jitter (m/s)\dotfill &$12^{+17}_{-13}$&$5.3^{+1.8}_{-3.5}$&$6.4^{+28}_{-6.4}$\\
~~~~$\sigma_J^2$\dotfill &RV Jitter Variance \dotfill &$170^{+750}_{-210}$&$27^{+22}_{-25}$&$40^{+1100}_{-590}$\\
\smallskip\\\multicolumn{2}{l}{Transit Parameters:}&Kepler UT 2015-10-08 (Kepler)\smallskip\\
~~~~$\sigma^{2}$\dotfill &Added Variance \dotfill &$1.972^{+0.061}_{-0.054} \times 10^{-8}$\\
~~~~$F_0$\dotfill &Baseline flux \dotfill &$0.9999990\pm0.0000023$\\
\enddata
\label{tab:cww89a}
\tablenotetext{}{See Table 3 in \citet{eastman2019} for a detailed description of all parameters}
\tablenotetext{1}{Time of conjunction is commonly reported as the ``transit time''}
\tablenotetext{2}{\bjdtdb is the target's barycentric frame and corrects for light travel time}
\tablenotetext{3}{Time of minimum projected separation is a more correct ``transit time''}
\tablenotetext{4}{At the epoch that minimizes the covariance between $T_C$ and Period}
\tablenotetext{5}{Use this to predict future transit times}
\tablenotetext{6}{Assumes no albedo and perfect redistribution}
\end{deluxetable*}

\end{document}